\documentclass{iaus}
\usepackage{graphicx}

\title[M51 Star Clusters] 
{Star Cluster Population of the Interacting Galaxy System M51}

\author[Hwang \& Lee]   
{Narae Hwang$^1$ \and Myung Gyoon Lee$^1$}%

\affiliation{$^1$Astronomy Program, Department of Physics and Astronomy, Seoul National University,
\break Seoul 151-742, Korea \break email: nhwang@astro.snu.ac.kr, mglee@astrog.snu.ac.kr}

\pubyear{2007}
\volume{241}  
\pagerange{119--126}
\date{?? and in revised form ??}
\jname{Stellar Populations as Building Blocks of Galaxies}
\editors{A. Vazdekis, \& R. Peletier, eds.}
\begin{document}

\maketitle

\begin{abstract}
We present a star cluster population study in the interacting galaxy system M51 based on HST ACS $BVI$ mosaic images taken by the Hubble Heritage Team to commemorate the HST's 15th anniversary. We have found and classified star clusters in M51 using SExtractor and visual inspection. We have derived the photometry, size, and age of the clusters. It is found that the companion SB0 galaxy NGC 5195 harbors about 50 faint fuzzy clusters
and that the age distribution of star clusters appears to be correlated with the epochs of dynamical events in M51 system.

\end{abstract}

\firstsection 
\section{Star Cluster Survey in M51}

Star clusters are an excellent tracer for stellar populations in the galaxy.
Being a nearby spiral galaxy with very low inclination,
M51 is one of the best targets for the star cluster studies in the late type galaxy. 
Recently, the Hubble Heritage program provided M51 images in $B, V, I,$ and $H \alpha$ bands
taken with HST/ACS (Mutchler \etal\ 2005)\cite[]{mut05}
and a few early results of star cluster studies based on this data are already available
(e.g., Gieles \etal\ 2006\cite[]{gie06}; Hwang \& Lee 2006\cite[]{hwa06}; 
Scheepmaker \etal\ 2006\cite[]{sch06}; Haas \etal\ 2006).

We have detected star cluster candidates using SExtractor (Bertin \& Arnouts 1996)\cite[]{ber96}
from the HST/ACS data
and then visually inspected these candidates with $V < 23$ mag to select the clean sample of star clusters.
The final star cluster catalog includes about 2200 Class 1 (circular shape without neighbor) and 1400 Class 2
(non-circular shape and/or with neighbors) star clusters.
Detailed information on the source detection and star cluster classification will be provided in \cite{hwa07}.
Star clusters in M51 are mostly distributed along the spiral structures of the galaxy.
However, Class 2 clusters are more tightly bound along the spiral arms while some Class 1 clusters are
even found in the inter-arm regions.
Comparison with the HII region catalog (Lee \etal\ 2007 {\it in preparation})
reveals that Class 2 clusters show stronger associations with HII regions than Class 1 clusters.
Most star clusters in M51, regardless of Class 1 or 2, are bluer than $(V-I) = 1.0$
but some clusters are as red as $(V-I) >1.0$ and these clusters tend to be found farther than $200''$
from the M51 center.

\begin{figure}[ht]
\begin{minipage}[t]{6.5cm}
 \begin{center}
 \includegraphics[width=5cm,clip]{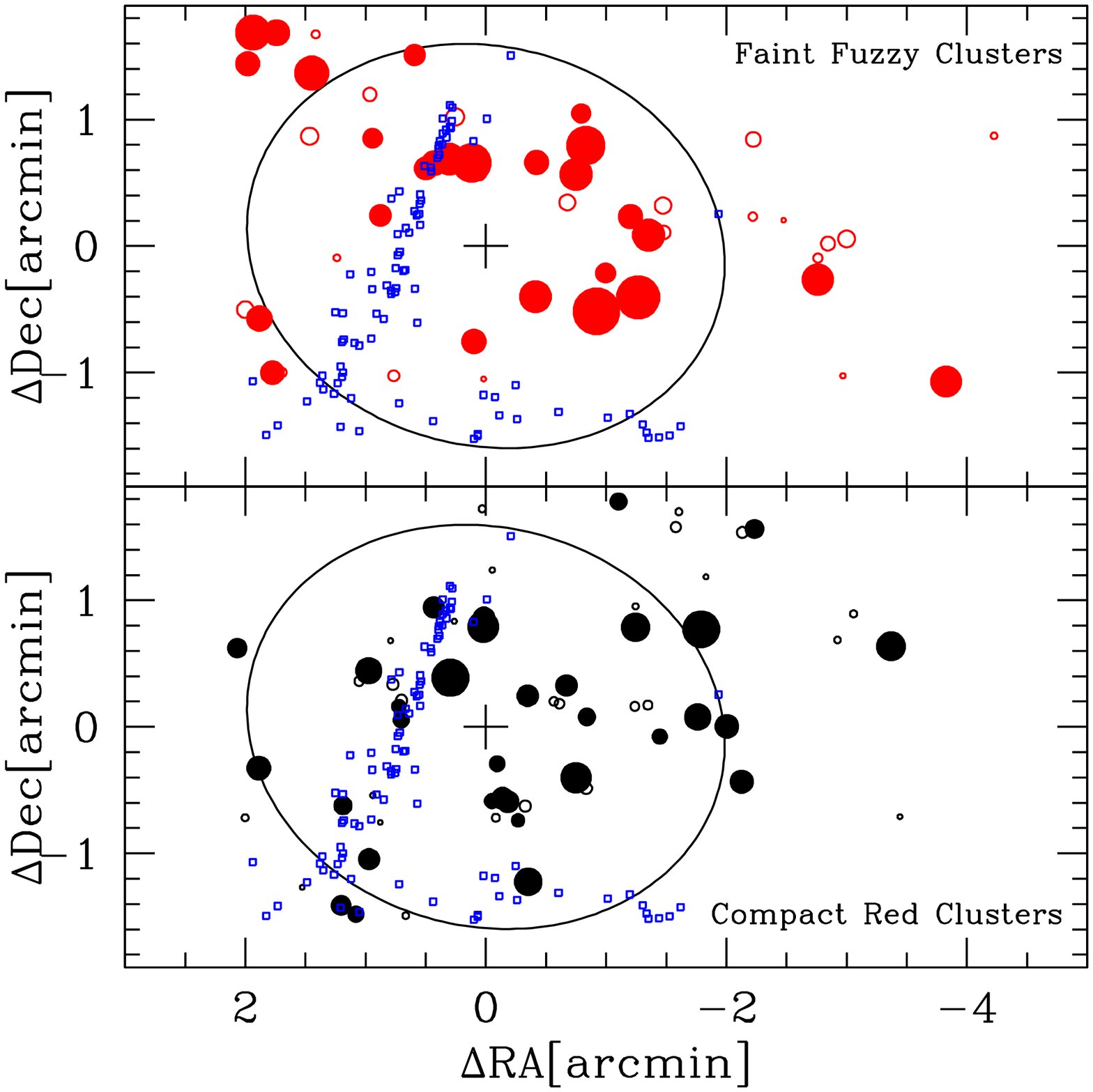}
 \caption{The spatial distribution of faint fuzzy clusters (upper) and `compact red' clusters (lower)
in NGC 5195. In each panel, clusters with $V<23.3$ are marked by filled circles and faint clusters with
$V \geq 23.3$ by open circles. The size of the circles is proportional to the luminosity of the corresponding
clusters: the bigger, the brighter.}
 \end{center}
\end{minipage}
\hfill
\begin{minipage}[t]{6.5cm}
 \begin{center}
  \includegraphics[width=5cm,clip]{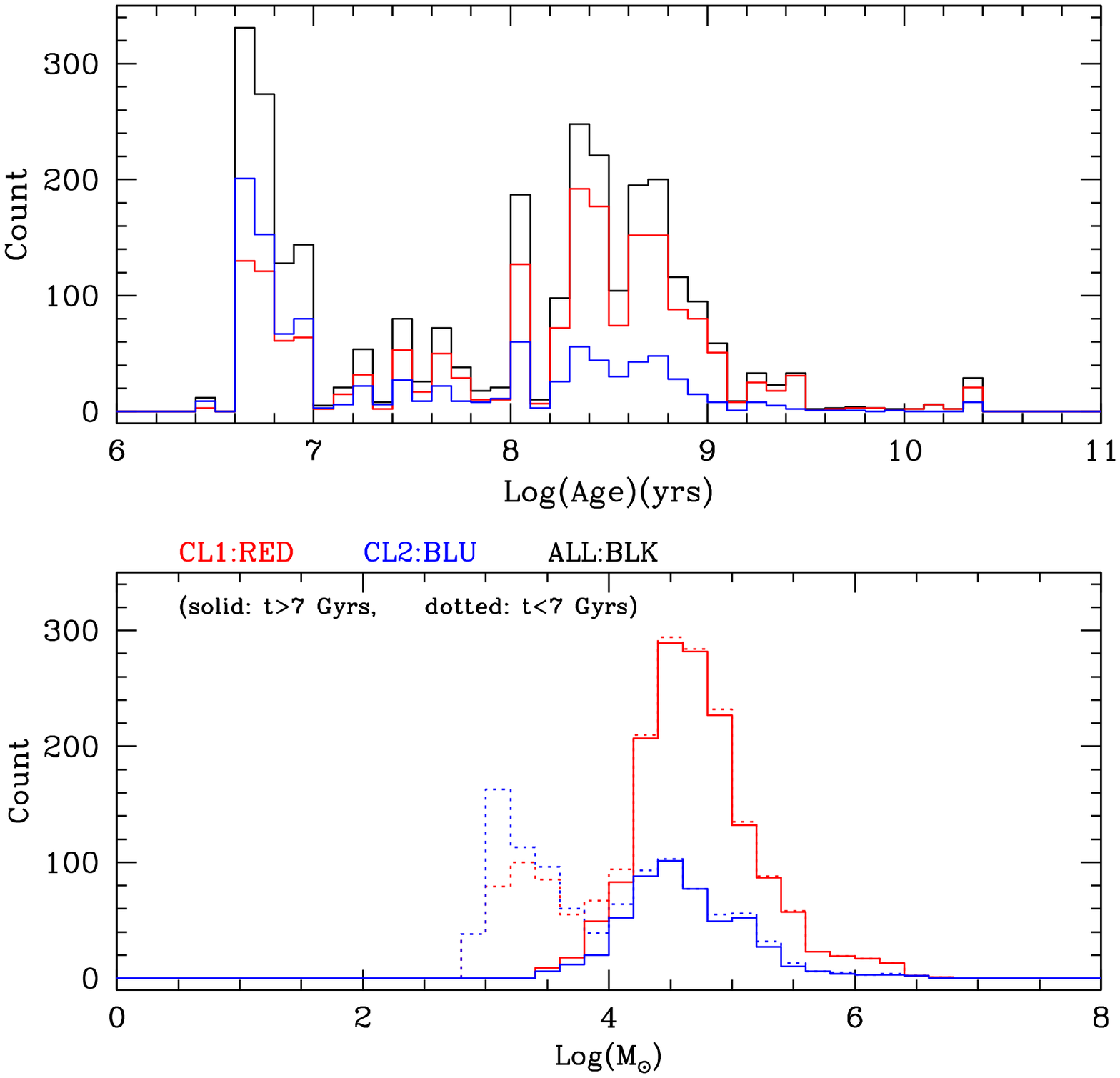}
  \caption{The age and mass distribution of star clusters in M51.
There are three prominent age peaks at about 100, 250, and 500 Myrs.
The mass distribution shows that most star clusters are as massive as $10^4 \sim 10^6$ $\rm{M}_{\odot}$.
The large age peak at $<10^7$ yrs appears not to be a real feature but it still represents the existence of very young star clusters.}
 \end{center}
\end{minipage}
\end{figure}

\section{Size and Age of Star Clusters}

We have measured the size of all star clusters using ISHAPE (Larsen 1999)\cite[]{lar99}.
The result shows that most star clusters are as large as $r_{eff} \approx 3$ pc but there are also
significant number of clusters with $r_{eff} > 7$ pc.
Many of these large clusters are also found to be redder than $(V-I) = 1.0$, which makes them
`faint fuzzy' clusters.
Interestingly, the faint fuzzy clusters are mostly located around the SB0 galaxy NGC 5195
and they show different spatial distribution from the red [$(V-I)>1.0$] and compact [$r_{eff}<7$ pc] clusters
as shown in Figure 1.
For furhter information, see \cite{hwa06}.

We have fitted the $BVI$ photometry data of star clusters to the model SED of \cite{bc03}
to estimate ages of star clusters.
Detailed information on the SED fitting will be given in our forthcoming paper.
Figure 2 shows the age distribution of star clusters and there are three peaks at about 100, 250, and 500 Myrs.
These ages roughly coincide with the encounter times of NGC 5194 and 5195
expected by theoretical models:
the single passage model (Toomre \& Toomre 1972\cite[]{too72}) expects the passage of NGC 5195
to have taken place at about $300 - 500$ Myrs ago while the multiple passage model
(Salo \& Laurikainen 2000\cite[]{sal00}) expects one more passage at about $50 - 100$ Myrs ago.
This result suggests that the dynamical history of the host galaxy is strongly correlated with the formation of star clusters.

\begin{acknowledgments}
N.H. is in part supported by the BK21 program of the Korean Government.
N.H. is grateful to IAU for its generous support through the IAU grant.

\end{acknowledgments}


\begin{thebibliography}{}

\bibitem[Bertin \&  Arnouts (1996)]{ber96} {Bertin, E., \& Arnounts, S.} 1996, \textit{A\&AS}, 117, 393

\bibitem[Bruzual \& Charlot (2003)]{bc03} {Bruzual, G., \& Charlot, S.} 2003, \textit{MNRAS}, 344, 1000

\bibitem[Gieles \etal\ (2006)]{gie06}
     {Gieles, M., \etal\ } 2006,
     \textit{A\&A}, 446, L9

\bibitem[Haas \etal\ (2006)]{haa06} 
     {Haas, M. R., \etal\ } 2006,
     \textit{astro-ph}/0609018

\bibitem[Hwang \& Lee (2006)]{hwa06}
     {Hwang, N., \& Lee, M. G.} 2006,
     \textit{ApJ}, 638, L79

\bibitem[Hwang \& Lee (2007)]{hwa07}
     {Hwang, N., \& Lee, M. G.} 2007,
     \textit{to be submitted}

\bibitem[Larsen (1999)]{lar99} {Larsen, S. S.} 1999, \textit{A\&AS}, 139, 393

\bibitem[Mutchler \etal\ (2005)]{mut05} {Mutchler, M. \etal} 2005, \textit{BAAS}, 37, 2


\bibitem[Salo \& Laurikainen (2000)]{sal00} {Salo, H., \& Laurikainen, E.} 2000a,
     \textit{MNRAS}, 319, 377

\bibitem[Scheepmaker \etal\ (2006)]{sch06} 
     {Scheepmaker, R. A., \etal\ } 2006,
     \textit{astro-ph}/0605022

\bibitem[Toomre \& Toomre (1972)]{too72} {Toomre, R. P. J., \& Toomre, J.} 1972, 
     \textit{ApJ}, 178, 623

\end{thebibliography}
\end{document}